\documentclass[twocolumn,showpacs,aps,floatfix]{revtex4}
\usepackage{graphicx}
\usepackage{bm}
\usepackage{amsmath}
\newcommand{\ignore}[1]{}
\def\half{{1 \over 2}}

\def\vt{\vartheta}

\def\cJ{{\cal J}} 
\def\ra{\rangle}
\def\la{\langle}

\begin{document}
%
\title{Memory effects induced by initial switching conditions}  
\author{J. Martorell}
\email{martorell@ecm.ub.es}
\affiliation{Departament d'Estructura i Constituents de la Materia, 
Facultat F\'{\i}sica, \\ 
University of Barcelona, Barcelona 08028, Spain}
\author{D. W. L. Sprung and W. van Dijk}
\affiliation{
Department of Physics and Astronomy, McMaster University, \\
  Hamilton, Ontario L8S 4M1 Canada}
\author{J. G. Muga}
\email{jg.muga@ehu.es}
\affiliation{
Departamento de Qu\'\i mica-F\'\i sica, UPV-EHU, Apdo 644,   
  48080 Bilbao, Spain}
\pacs{03.65.Wj,   
      21.10.Tg,    
      42.50.Xa }   
\begin{abstract}
Initial-switching refers to the way in which the decay of an 
initially confined state begins, as the barrier isolating 
it from the exterior is relaxed. We study these effects in the context of 
Longhi's version of the Fano-Anderson model. Most authors  assume the 
sudden approximation where the coupling is turned on instantaneously. We 
consider a finite rise time $T$, both numerically and analytically. 
When the coupling is ramped up linearly over a switching time $T$, we show 
that the asymptotic survival amplitude acquires a phase $T$ and is 
modulated by a factor $(\sin T) / T$. Several other results relating to 
the solution of the model are obtained. All site amplitudes have the 
same decay constant during the exponential decay regime. In the 
asymptotic regime, the amplitude and phase of decay oscillations 
depend on the initial-switching profile, but the period does not.   
\end{abstract}
%
\maketitle
\section{Introduction}
Quantal effects in the time evolution of decaying systems have been 
studied since the beginning of quantum mechanics. Gamow's explanation 
of alpha-decay was among the first successful applications of 
quantum theory to radioactive nuclei~\cite{Ra03}. The theory showed 
that decay was exponential in the accessible range of times, and 
related the lifetime to the nuclear charge $Z$ and decay energy. 
Khalfin~\cite{Kh57} was the first to realize that, when the energy 
spectrum of the system is bounded from below, exponential decay 
cannot persist as $t \to \infty$. Eventually it has to be replaced by an 
asymptotic regime with a slower rate of decrease. Explicit solution 
of the Schr\"odinger equation for simple models of a particle 
escaping from a confining potential verified that in the asymptotic 
regime, the survival (or non-escape) probability shows power-law 
decrease~\cite{Winter61}. Under rather general assumptions, this 
power-law can be shown to be $\propto t^{-3}$ for finite range 
potentials~\cite{GW64}. 
Some additional aspects are discussed in~\cite{MMS08,MMS09}, and 
references therein. Exponential decay is observed in a wide variety 
of physical systems. The predicted slower decay rate at long times 
has proven difficult to verify, but it is a universal feature present 
in theoretical models. This applies not only to a particle trapped by  
a potential barrier, but also to the decay of a discrete state coupled to a 
continuum~\cite{CT92,NNP96}. 

It is less well-known that when the confining barrier, or the 
coupling between discrete state and continuum, is time-dependent in the 
initial stages of decay, so-called ``memory effects'' affect the 
post-exponential survival probability of the system. 
These effects were proposed as a means of enhancing the survival 
probability, to make it feasible to observe the post-exponential regime 
experimentally. In particular, Robinson~\cite{Rob86} considered 
adiabatic switching-on of the system over an infinite time period, to 
clarify the role of switching time versus observation time. 

Tight binding models have been widely used in condensed matter for many years
\cite{AM}, \cite{Ki} to describe tightly bound bands
in crystals. More recently they have been applied to  multilayered
semiconductor heterostructures \cite{Dav99}, to the dynamics of Bloch
oscillations \cite{Ha04}  and to the study of properties of Bose
condensates in a lattice \cite{ST03}. They are particularly well suited
when the unit cell of the periodic struture leads to a well confined lowest
energy state.  

In this paper, we 
examine memory effects for finite switching times. We use a 
variant of the Fano-Anderson model~\cite{Fa61,An61,Ma90} 
proposed by Longhi~\cite{Lo06} to study deviations from 
exponential decay at short times, and the Zeno effect~\cite{FP08}. 
His model can be viewed as the decay of a discrete state coupled to a 
single-band tight-binding continuum. But, more interestingly from an 
experimental point of view, he presented it as a one dimensional 
semi-infinite periodic chain of discrete sites. The coupling between 
adjacent sites, $g\, ( |n \ra \la n+1| + |n+1 \ra\la n| )$, is described by 
a universal hopping constant, $g$, and the 
site energies are taken all to be equal. For two sites isolated from 
the rest, $\pi \hbar/(2g)$ would be the half-period, the time for a 
complete transfer of occupation between the pair. 
Relative isolation of the first site 
is obtained by fixing its coupling parameter $g_1 \equiv g \Delta$ to 
be different from the rest. Under these conditions, when at $t=0$ only 
site 1 is occupied, Longhi obtained the survival probability, its 
lifetime for exponential decay and the asymptotic power-
law~\cite{Lo06,Lo06a}. The connection to models of a confined 
particle, like alpha-decay, views site 1 as the well, the ratio 
$\Delta = g_1/g$ as the effect of a confining barrier, and the semi-
infinite chain of sites in position space as a single-band continuum.
 \begin{figure} [htb]                         
\includegraphics[width=8cm]{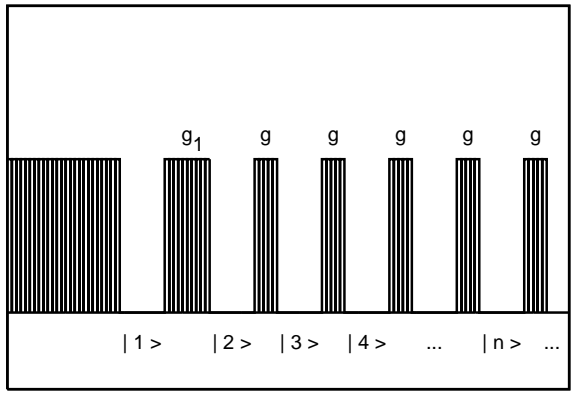}
\caption{Semi-infinite array of identical 1D wells coupled to a ``defect''
  well separated by a thicker barrier.} 
 \label{tb1x}
\end{figure} 
Fig. \ref{tb1x} illustrates the system we have in mind. The barrier 
separating site 1 from the rest of a semi-infinite array is thicker 
than the others, to reduce tunnelling,
which in a tight binding model implies $g_1 < g$. A thick enough barrier
effectively makes $g_1 = 0$, removing site 1 from consideration.

\begin{figure} [htb]                         
 \includegraphics[width=8cm]{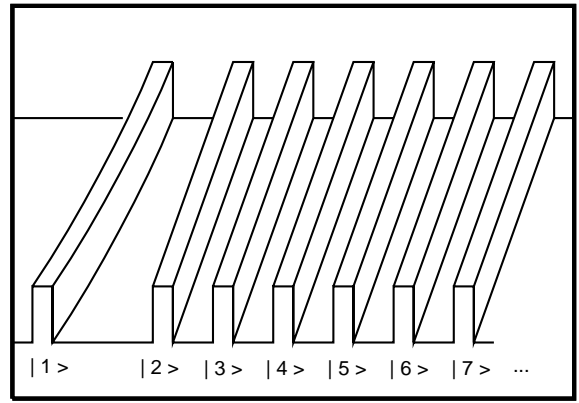}
\caption{A segment of a semi-infinite array of tunnel-coupled optical 
waveguides, made by etching. The $z$-axis is parallel to the guides, 
and $z_T$ at the TOP of the drawing. A logarithmic dependence on $z$ 
of the separation leads to linear dependence of the coupling.  } 
 \label{optw2}
\end{figure}

   An added interest of the model is that it has a classical 
electro-magnetic analogue, in the form of a waveguide 
array~\cite{Jo65, CLS03}. Fig. \ref{optw2} describes one such semi-infinte
array with an initial ``defect''. Fig. 1(a) of \cite{CLS03} shows an
experimental realization of such an array. 
Consider an ideal semi-infinite array of 
long parallel waveguides. The variation with $z$, (the longitudinal 
distance along the guide) of the electric field in the $n$'th guide, 
replaces the variation with time of the site amplitudes. Both systems 
obey a set of coupled mode equations formally the same as those for 
the time variation of the site amplitudes in the Fano-Anderson model. 
Recent experiments, using scanning tunnelling microscopy with 
sensitive near-field probes, have allowed Longhi and 
collaborators~\cite{VLL07} to measure the evanescent fields along the 
waveguides, and provide a convincing quantitative demonstration of 
the classical-quantal analogy for the case of a periodic parallel 
array. In addition the classical analogue of the quantum 
Zeno effect was verified~\cite{Bia08}. 
The validity of the Fano-Anderson model to describe such effects has been
quantitatively shown by these authors~\cite{VLL07,Bia08}.

In the present work, we study the problem of initial switching, 
by letting $g_1(t)/g \equiv \Delta_v(t)$ depend on time, during a 
finite initial time period $0< t < T$, following which $\Delta$  is 
constant. 
In Longhi's optical analogue, this could be implemented by having many   
parallel waveguides coupled to a first curved waveguide, with its
separation distance decreasing smoothly with $z$ from  an initially 
large value (where tunnelling is practically negligible), to a 
constant value beyond $z=z_T$. The final separation distance should be 
chosen to give well-expressed exponential decay. Longhi {\it et 
al}.~\cite{LDOL07}
have demonstrated the experimental feasibility  of a similar device 
(see their Fig. 1). Our work differs from that of Robinson~\cite{Rob86}, who
considered a smooth switching function increasing adiabatically over 
$-\infty < t < \infty$, so his $\Delta_R(t)$ saturates only asymptotically.  
Finite switching times have also been considered in the context of threshold 
effects in pulsed laser photionization; see Chapter 6 of~\cite{RZ93} for a 
review. 
In Section II we discuss solutions of Longhi's model in the exponential 
and asymptotic regimes.
In Section III  we derive analytic expressions for 
the time evolution during the period of varying coupling strength. 
Specific calculations of the effects are reported in Section IV. 
Verification of these predictions should be possible with present  
experimental capabilities. Verification of the predicted memory 
effects at long times may require further development of experimental 
techniques.

\section{Decay for arbitrary initial conditions}
\begin{figure} [htb]                         
 \includegraphics[width=8cm]{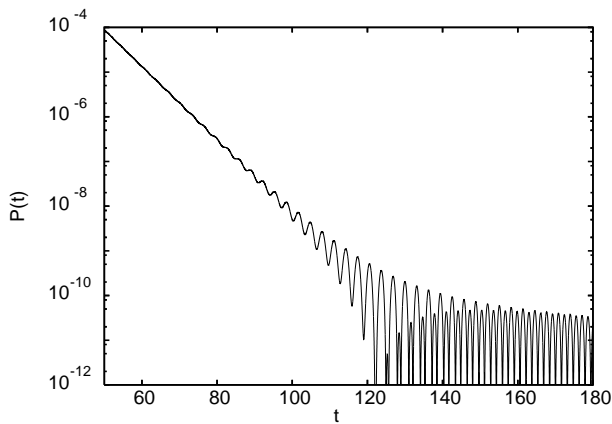}
\caption{Survival probability $P(t)$, for constant $\Delta = 0.3$, in the 
region of changeover from exponential to asymptotic decay.  Here and in the
subsequent  figures, time $t$ ( dimensionless) is in units $\hbar/g$.} 
 \label{f1pt03}
\end{figure}

To begin, we study the time evolution of a semi-infinite tight 
binding system that initially has a single site $n=q$ occupied at $t=0$.
For convenience, we will use dimensionless units. Energies will be 
measured in units of $g$ (the hopping parameter of the periodic 
lattice), and time in units of $\hbar/g$. Wth the site energies set to 
zero, the  Hamiltonian is 
 \begin{eqnarray}                      
H &=& -\Delta_v(t) \bigg[ |1 \ra\la 2| + |2 \ra\la 1|\bigg]  \nonumber \\ 
&-&  \sum_{n=2}^{\infty}\, \bigg[ |n \ra\la n+1|+ |n+1 \ra\la n| \bigg].
\label{eq:idr01}
\end{eqnarray}
We assume a positive coupling parameter  $0 <  \Delta <1$. 
Longhi~\cite{Lo06,Lo06a} derived analytic expressions for the 
occupation amplitude $c_n(t)$, focussed primarily on the case that 
only the first site $q=1$ is occupied at time zero, and constant $\Delta$. 
Fig. \ref{f1pt03}  shows that the choice $\Delta =0.3$ gives a 
survival probability, $P(t) = |c_1(t)|^2$, which decays exponentially 
for a considerable range of times, and  afterward the 
post-exponential decay consists of oscillations whose envelope follows the 
$t^{-3}$ law. This will be our example of reference in the remainder 
of the article. 

In this paper we suppose that during an initial period $0 < t \le T$, 
the coupling parameter $\Delta_v$ increases continuously from $0$ 
to $\Delta$ and afterwards remains constant. In Section III we will 
study a specific choice for the initial time-dependence, that 
we call linear rise 
 \begin{eqnarray}                      
\Delta_v(t)= \Delta_\ell(t) &\equiv& t \ \Delta/T~, \quad t \le T \nonumber  
\\ 
&=& \Delta~, \qquad \quad t > T.
\label{eq:idr02}
\end{eqnarray}
For this model we will derive explicit expressions for the survival 
amplitude and other quantities, valid to lowest order in $\Delta$. 
Exponential decay over a sizeable time scale requires that $\Delta$ 
be small compared to 1, so linear terms should suffice to establish 
the importance of such effects.  

Initially we focus on general properties that apply to any 
scheme for the initial switching-on of $\Delta$. Given {\it any} profile  
$\Delta_v(t)$, the site occupation amplitudes, $c_n(t)$, $n = 1, 2, \cdots$ 
can be determined by numerical solution of the time-dependent 
Schr\"odinger equation (TDSE), for which we use the method described 
in~\cite{Koo90}. Alternatively, in Sec. III we use the time evolution 
operator to derive approximate analytic results. In either approach, 
the end result is a set of occupation amplitudes at $t=T$ which in 
this Section we assume to be known, and that we write in vector 
form 
 \begin{equation}                       
|{\vec \vt} \ra = \begin{pmatrix} \vt_1 &  \vt_2 &  .. & \cdots & \vt_q &
 \cdots \end{pmatrix}^{\mathsf{T}}
\label{eq:idr03}
\end{equation}
with $\sum_{q=1}^{\infty} |\vt_q|^2 = 1$ and where by ${\mathsf{T}}$ we
denote the transpose. Since we will always assume 
that after time $T$,  $\Delta_v(t\ge T)= \Delta$ remains constant, it is 
convenient to reset the zero of time by writing ${\tilde  t}=t-T$. 
In the remainder of this section we study the time evolution of 
a system with these evolved initial conditions, at ${\tilde t}=0$,  given by 
Eq. (\ref{eq:idr03}), and constant $\Delta$. The further evolution is governed 
by the TDSE 
 \begin{equation}                       
i \frac{\partial}{\partial {\tilde  t}}  |{\vec \psi}({\tilde  t}) \ra
= H |{\vec \psi}({\tilde  t}) \ra 
\label{eq:idr04}
\end{equation}
with
 \begin{equation}                       
|{\vec \psi}({\tilde t}) \ra = (c_1({\tilde t}), c_2({\tilde t}), ..,
 c_n({\tilde t}), .. )^{\mathsf{T}}.
\label{eq:idr05}
\end{equation}
The linearity in time of the  TDSE implies that 
 \begin{equation}                       
c_n({\tilde  t}) = \sum_{q=1}^{\infty} \vt_q \ c_{n,q}({\tilde  t})  
\label{eq:idr06}
\end{equation}
where the $c_{n,q}({\tilde  t})$ are the site amplitudes of a state, 
$|{\vec \psi}_q({\tilde  t}) \ra$, which at ${\tilde  t}=0$ has unit real    
amplitude $c_{q,q}(0) = 1$  on the $q$-th site, and vanishes on all others.
 \begin{eqnarray}                       
|{\vec \psi}_q({\tilde  t}=0) \ra^{\mathsf{T}} &=& ( 0, 0, \cdots, 1, 0, 
\cdots ) 
\nonumber \\
|{\vec \psi}_q({\tilde  t}) \ra^{\mathsf{T}} &=& (c_{1,q}({\tilde  t}), c_{2,q}
({\tilde  t}) , \cdots c_{n,q}({\tilde  t}), \cdots ) 
\label{eq:idr07}
\end{eqnarray}  
\paragraph*{Stationary solutions.}
To determine these $c_{n, q}({\tilde  t})$, we first find stationary 
solutions of the coupled equations, and then propagate linear 
combinations of them in time.   Writing $|{\vec \psi}({\tilde  t}) 
\ra =$ exp$(- iE{\tilde  t}) |{\vec u} \ra$, the components  of 
$|{\vec u} \ra$ satisfy 
 \begin{eqnarray}                       
0 & = & 2\cos \phi\, u_1 -\Delta\, u_2 \nonumber \\
0&=& -\Delta\, u_1 + 2\cos \phi\, u_2 -u_3 \nonumber \\
0&=& -u_{n-1} + 2\cos \phi\, u_n -u_{n+1} \quad , \quad n \ge 3 
\label{eq:idr08}
\end{eqnarray}
where we have written $E \equiv -2 \cos \phi$. 
The solutio is most easily obtained by expressing $u_n$  in terms of
incoming and outgoing waves with amplitudes denoted ${\cal I}$ and ${\cal O}$: 
 \begin{eqnarray}                       
u_n &=& {\cal I} e^{i(n-2)\phi} -{\cal O} e^{-i(n-2)\phi} \quad , n\ge 2
\nonumber \\
u_1 &=& \Delta \big({\cal I}-{\cal O}\big) /(2~\cos \phi) \nonumber \\
\Delta u_1 &=& 2~\cos \phi\ u_2-u_3 .
\label{eq:idr09}
\end{eqnarray}
Inserting the first and second lines into the third, gives 
 \begin{equation}                       
\frac{{\cal O}}{{\cal I}} = \frac{2\cos \phi~e^{-i\phi} -\Delta^2}
 {2\cos \phi~e^{i\phi}  -\Delta^2} 
= \frac{\alpha^2 + e^{-2i\phi}}{\alpha^2 + e^{+2i\phi}}
\equiv r(\phi)~,
\label{eq:idr10}
\end{equation}
where $\alpha^2 \equiv 1- \Delta^2$ is  the decoupling parameter.
Since the normalization of the states $|{\vec u} \ra$  is arbitrary at 
this stage, we may define the amplitudes ${\cal O}$ and ${\cal I}$ as the 
numerator and denominator of Eq. (\ref{eq:idr10}). Then the  
$u_n $ in Eq. (\ref{eq:idr09}) take the compact form 
 \begin{eqnarray}                       
u_n(\phi) &=& 2i \bigg( \sin n \phi +\alpha^2 \sin (n-2)\phi \bigg)~, 
\quad  n\ge 2 \nonumber \\
u_1(\phi) &=& 2i \Delta\sin \phi~, 
\label{eq:idr11}
\end{eqnarray}
(Notice that if line one is 
specialized to $n=1$, we obtain $\Delta$ times $u_1(\phi)$. Therefore 
one can avoid treating $n=1$ as a special case, if we simply remember 
to remove this extra factor.) 
\paragraph*{Time dependent solutions:}
Following~\cite{Lo06a} we write the $c_n({\tilde  t})$
as weighted integrals over the stationary solutions, with a weight function
$F(\phi)$ (still to be determined) expressed as a Fourier series
 \begin{eqnarray}                       
F(\phi) &\equiv& \frac{1}{2\pi \Delta} \sum_{N=1}^{\infty} \ T_N\
e^{-iN\phi}~;  \nonumber \\
c_n({\tilde  t} )&=& \int_{-\pi}^{\pi} \ d\phi\ u_n(\phi) \ 
e^{-iE{\tilde  t}}\ F(\phi)~.  
\label{eq:idr12}
\end{eqnarray}
The parameters $T_N$ are determined by the initial conditions, as we now 
explain. In what follows we will often use properties of Bessel 
functions which can be found in Chapter 11 of Arfken~\cite{Ar85}; to 
avoid tiresome repetition, we will cite it no further. 
For convenience we introduce symbols $\cJ_q(2t) \equiv i^q J_q(2t)$, 
which have the symmetry $\cJ_q = \cJ_{-q}$, and satisfy the 
recurrence relation  
 \begin{equation}                       
\cJ_{n-1} - \cJ_{n+1} =  \frac{n}{i {t}} \cJ_n. 
\label{eq:idr13}
\end{equation}
This allows us to write 
 \begin{eqnarray}                       
I_{p,N} &=& \frac{i}{\pi} \int_{-\pi}^{\pi} \ d\phi \ \sin p \phi \
e^{i(2{\tilde  t}\cos \phi- N\phi)}  \nonumber \\
&=& i^{N-p}  J_{N-p}(2{\tilde  t}) -i^{N+p} J_{N+p}(2{\tilde  t})\nonumber \\
&\equiv&  \cJ_{N-p}(2{\tilde  t}) -\cJ_{N+p}(2{\tilde  t})  
\label{eq:idr14}  \\  
&=&  \cJ_{N-p} -\cJ_{N-p+2} + \cJ_{N-p+2} -\cJ_{N-p+4} + \nonumber \\
&& \quad \cdots + \cJ_{N+p-2}-\cJ_{N+p}\nonumber \\
&=& \sum_{m=1}^p \frac{(N+p+1-2m)}{i{\tilde  t}} \cJ_{N+p+1-2m} \ .
\label{eq:idr15}
\end{eqnarray}
In the limit of large argument $2 t \gg N+p$, $I_{p,N}$  approaches its 
asymptotic form  
\begin{eqnarray}                        
 I_{p,N}^{(a)} &\equiv& \sum_{m=1}^p \frac{ (N+p+1-2m)}{\sqrt{\pi} 
{\tilde  t}^{3/2}} i^{N+p-2m}\nonumber \\
&& \qquad  \cos[ 2{\tilde  t} - (N+p+1-2m +1/2)\pi/2] \nonumber \\
&=& \sum_{m=1}^p \frac{(N+p+1-2m)}{\sqrt{\pi} {\tilde  t}^{3/2}} 
i^{N+p} \times \nonumber \\
&& \qquad \quad \cos[ 2{\tilde  t} - (N+p+1 +1/2)\pi/2]\nonumber \\
&=&  \frac{i^{N+p-2} }{\sqrt{\pi}~{\tilde  t}^{3/2}} \sum_{m=1}^p (N+p+1-2m)
\times \nonumber \\
&& \qquad \quad \cos[ 2{\tilde  t} - (N+p-1/2)\pi/2]  \nonumber \\
&=&   \frac{i^{N+p-2} }{\sqrt{\pi}~{\tilde  t}^{3/2}}\, N~p \,
\cos\left[ 2{\tilde  t} - \left(N+p-\half\right)\frac{\pi}{2}\right].
\label{eq:idr16}
\end{eqnarray}
Using Eqs. (\ref{eq:idr11}), (\ref{eq:idr12}) and (\ref{eq:idr14}), 
\begin{eqnarray}                       
c_n({\tilde  t}) &=& \frac{1}{\Delta} \sum_{N=1} T_N 
\bigg[ (\cJ_{N-n}( 2{\tilde  t} ) -\cJ_{N+n}( 2{\tilde  t} ) )
  \nonumber \\
&& \,\,\, + \alpha^2 \ \left( \cJ_{N-n+2}( 2{\tilde  t} ) -  \cJ_{N+n-2}( 
2{\tilde  t} ) \right) \bigg]; \nonumber \\ 
&& \qquad \qquad  \qquad n \ge 2~,  \qquad {\rm while} \nonumber \\
c_1({\tilde  t}) &=& \sum_{N=1} T_N \ \big( \cJ_{N-1}( 2{\tilde  t} ) 
 - \cJ_{N+1}( 2{\tilde  t} ) \big) \nonumber \\
&=& \sum_{N=1} T_N \ \frac{N}{i \tilde  t} \cJ_N( 2{\tilde  t} ) \nonumber \\ 
&=&  \sum_{N=1} T_N \ i^{N-1} \frac{N}{\tilde  t} J_N( 2{\tilde  t} ),
\label{eq:idr17}
\end{eqnarray}
so that at ${\tilde  t}=0$:
 \begin{eqnarray}                       
\Delta\,\, c_n(0) &=&  (T_n + \alpha^2 T_{n-2})  \quad , \quad  n \ge 2
\nonumber \\
c_1(0) &=& T_1.
\label{eq:idr18}
\end{eqnarray}
Note that $T_0=0$, since only the sites with positive indices exist. 
Given initial conditions for the $c_n({\tilde t})$, as e.g. in 
Eq. (\ref{eq:idr03}), one determines the $T_N,\ N=1, 2, ..$ from the  
recurrence relation Eq. (\ref{eq:idr18}). This completes construction 
of the time-dependent solution. 

Up to now, our results are valid for all times ${\tilde t} > 0 $. In the 
next two subsections we develop approximations valid in the 
asymptotic and exponential regimes which are illustrated in Fig. 1. 

\subsection{Asymptotic approximation} 

Since $J_N(2{\tilde t})/{\tilde t} \propto ( {\tilde t} )^{-3/2}$ when 
${\tilde t} \to \infty$,  one sees that for long times $|c_1({\tilde t})|^2 
\propto {\tilde t}^{-3}$ for any set of initial occupations, as  
stated in the Introduction. Using Eqs. (\ref{eq:idr16}) and 
(\ref{eq:idr17}) we find 
\begin{eqnarray}                       
&& c_1({\tilde t}) \simeq c_1^{(a)}({\tilde t}) \nonumber \\ &=&
\frac{1}{\sqrt{\pi} {\tilde t}^{3/2}} \sum_{N=1} N T_N \ i^{N-1} \cos
\left(2{\tilde t}- N\frac{\pi}{2} - \frac{\pi}{4} \right) \nonumber \\
&=& \frac{1}{\sqrt{\pi} {\tilde t}^{3/2}} \bigg[\left(\sum_{l=0} (2l+1) 
T_{2l+1}\right) \sin\left(2{\tilde t}-\frac{\pi}{4}\right) \nonumber \\ 
&&\quad - i \left( \sum_{l=0}\ 2l\ T_{2l} \right) \cos\left(2{\tilde t} 
- \frac{\pi}{4}\right) \bigg], 
\label{eq:idr19}
\end{eqnarray}
where for later convenience we have isolated the contributions from 
even and odd order $T_N$. Providing that the $T_N$ are real, 
 (which they will be,) 
this is equivalent to separating the real and imaginary parts.

For the asymptotic behaviour of the $c_n$, $ n \ge 2$,  we use 
Eqs. (\ref{eq:idr14}), (\ref{eq:idr16}) and (\ref{eq:idr17}) 
to arrive at 
\begin{eqnarray}                       
&& c_n^{(a)}({\tilde t}) =  \frac{i^{n-1}}{\sqrt{\pi}\ {\tilde 
t}^{3/2}\ \Delta} \bigg(n+ \alpha^2(n-2) \bigg) \times \nonumber \\ 
&& \,\,\sum_{N=1}  i^{N-1} \ N \ T_N  \cos\left(2{\tilde t}
  -\left(N+n-\frac{1}{2}\right)\frac{\pi}{2}\right)  \nonumber \\ 
&& \,\, = \frac{i^{n-1} \bigg(n+ \alpha^2(n-2) \bigg)}{\sqrt{\pi}\,
{\tilde t}^{3/2}\  \Delta}   \times \nonumber \\ 
&&\,\,\bigg[  \left(\sum_{l=0} (2l+1) T_{2l+1} \right) \cos \left(2{\tilde t}
-(n+1/2)\frac{\pi}{2}\right)   \nonumber \\  
&&\quad  
- i \left( \sum_{l=0} \ 2l \ T_{2l}\right)  \cos\left(2{\tilde t} 
- (n-1/2)\frac{\pi}{2}\right) \bigg], 
\label{eq:idr20}
\end{eqnarray}
for $n \ge 2$. Providing that $2{\tilde t} >> n\pi$, we can replace 
the cosines in the last lines by 
$J_n(2 {\tilde t})$ and $J_{n-1}(2 {\tilde t})$ respectively, 
(making appropriate adjustments to the prefactor). 

Now we particularise to cases where the initial amplitude is confined to 
a single site labelled $q$.  By combining these results, we can finally 
write the general solution for arbitrary initial occupations.
%
\paragraph*{Case 1: Initial amplitude only on site 1.}
At ${\tilde t}=0$ : $c_1(0)=1,\ c_n(0)=0,\ n 
\ge 2$. It is then immediate from Eq. (\ref{eq:idr18}) that $T_{2l}= 0$ 
for all $l$ and that $T_{2l+1}= (-)^l \alpha^{2l}$. Inserting these 
in Eq. (\ref{eq:idr19}) one finds 
 \begin{eqnarray}                       
c_{1,1}^{(a)} 
({\tilde t }) &=&  \frac{1}{\sqrt{\pi} {\tilde t 
}^{3/2}} \sin\left(2{\tilde t}-\frac{\pi}{4}\right) \sum_{l=0}
(2l+1)(-)^l \alpha^{2l} \nonumber \\
&=& \frac{1}{\sqrt{\pi} {\tilde t 
}^{3/2}} \ \frac{1-\alpha^2}{(1+\alpha^2)^2}
\sin\left(2{\tilde t}-\frac{\pi}{4}\right)
\label{eq:idr21}
\end{eqnarray}
in agreement with Eq. 8 of~\cite{Lo06}. Hereafter we will call this 
case the ``sudden approximation" because it corresponds to the abrupt 
switching-on of the coupling from $\Delta_v= 0$ to $\Delta_v 
= \Delta$ at $t=T=0$. 
\paragraph*{Case 2: Initial amplitude on another odd site.}
Consider now $c_n(0)=\delta_{n,q}$ , $q>1$ odd. Then $T_{2l} =0, \ 
l=0,1, ..$ and  
 \begin{eqnarray}                       
T_1 &=& T_3 = \cdots =T_{q-2}=0 \quad , \quad T_q= \Delta \nonumber \\
T_{q+2l} &=& (-)^l \alpha^{2l} \Delta~,
\label{eq:idr22}
\end{eqnarray} 
so that Eq. (\ref{eq:idr19}) gives 
 \begin{eqnarray}                       
c_{1,q}^{(a)}({\tilde  t}) 
&=&  \frac{\Delta}{\sqrt{\pi}\ {\tilde  t}^{3/2}}
\frac{q(1+\alpha^2)-2\alpha^2}{(1+\alpha^2)^2} 
\sin\left(2{\tilde  t}-\frac{\pi}{4}\right) .
\label{eq:idr23}
\end{eqnarray}
Note that for $q=1$ this result agrees with Eq. (\ref{eq:idr21}), if 
we remove  the explicit $\Delta$, in agreement with our earlier 
remark.  The meaning of the factor $\Delta$ is this: When site 1 is 
initially empty, the wave must cross the barrier at least once before 
its decay may begin. 
\paragraph*{Case  3: Initial amplitude on an even site.}
When $c_n(0)=\delta_{n,q}$, $q$ even,  we find $T_{2l+1} 
=0,\  l = 0, 1, \cdots$ and 
 \begin{eqnarray}                       
T_0 &=& T_2 = \cdots=T_{q-2}=0 \quad , \quad T_q= \Delta\nonumber \\
T_{q+2l} &=&(-)^l \alpha^{2l} \Delta.
\label{eq:idr24}
\end{eqnarray} 
From Eq. (\ref{eq:idr19}) 
 \begin{eqnarray}                       
c_{1,q}^{(a)} ({\tilde  t})
&=& \frac{-i\Delta}{\sqrt{\pi}\ {\tilde  t}^{3/2}}
\frac{q(1+\alpha^2)-2\alpha^2}{(1+\alpha^2)^2} \cos\left(2{\tilde
    t}-\frac{\pi}{4}\right).
\label{eq:idr25}
\end{eqnarray}
\paragraph*{General initial conditions.}
We now combine the previous results to obtain the most general 
solution. Inserting Eqs. (\ref{eq:idr23}) and (\ref{eq:idr25}) into 
(\ref{eq:idr06}) we find 
 \begin{eqnarray}                       
& &c_1^{(a)}({\tilde  t}) = \frac{1}{\sqrt{\pi} (1+\alpha^2)^2 \
  {\tilde  t}^{3/2}} \nonumber \\ && \quad 
\times \bigg[ S_{o} \sin \left(2{\tilde  t} -\frac{\pi}{4}\right) - 
i S_{e} \cos\left(2{\tilde  t}-\frac{\pi}{4}\right) \bigg]
\label{eq:idr26}
\end{eqnarray}
with the amplitudes summed over odd (even) indices 
 \begin{eqnarray}                       
S_{o(e)} &\equiv& \sum_{q = odd (even)}^{\infty} {\tilde \vt}_q 
[q(1+\alpha^2) - 2\alpha^2], 
\label{eq:idr27}
\end{eqnarray}
where we have defined ${\tilde \vt}_1 = \vt_1$ and 
${\tilde \vt}_q = \vt_q \Delta $, $ q = 2, 3, \cdots$. 

In principle, the phases of the amplitudes $\vt_q$ at time zero are 
arbitrary, so that $S_0$ and $S_e$ are neither purely real or 
imaginary. Then $c_1^{(a)}$ of Eq. (\ref{eq:idr26}) is a complex 
number and will have a node only accidentally. But in the case we 
wish to discuss, it is assumed that at some time in the past the 
system was prepared with only site 1 occupied and $c_1(t=0) = 1$. 
During the switching on process, the system evolved to build up the 
amplitudes in sites $2,\, 3,\, \cdots$. In the special 
case of sudden switching, $T=0$, we can apply Eq. (\ref{eq:idr20}) with
${\tilde t}=t$ to see that the resulting phase of $c_q(t)$ 
is $i^{q-1}$, the quantities $T_N$ being real. Using a perturbation 
expansion, we will see that the same phases are generated by 
the time-evolution operator in Section III (see discussion following 
Eq. (\ref{eq:idr61})).   As a result, the 
amplitudes on odd-numbered sites are real and those on even numbered 
sites, pure imaginary. Hence $S_o$ and $X_e =-iS_e$ are both real, 
allowing us to write 
\begin{eqnarray}                       
c_1^{(a)} ({\tilde  t}) &= & \frac{1}{\sqrt{\pi} (1+\alpha^2)^2 \
  {\tilde  t}^{3/2}} \ {\cal A}
\sin \left( 2{\tilde  t} -\frac{\pi}{4} + \Phi\right)\nonumber \\
{\cal A} &=& \sqrt{S_{o}^2+ X_{e}^2} \nonumber \\
\Phi &=& \arctan (X_{e}/ S_{o}). 
\label{eq:idr28}
\end{eqnarray}
This simplification ensures that the survival probability will still 
have nodes, and the  period of the asymptotic oscillations is not 
changed by the initial switching process. Compared to the sudden 
approximation Eq. (\ref{eq:idr21}), there is an extra phase $\Phi$, 
and the asymptotic amplitude involves ${\cal A}$ in place of $\Delta$. 
To have $\Phi = 0$ would require that the even-sites are empty at 
${\tilde t} = 0$.

\subsection{Exponential decay regime} 
Again, we first consider the situation where at ${\tilde t}=0$, a 
single site $q$ is occupied, and use these as 
building blocks for the general result to be given at the end. 
Inserting the expressions Eqs. (\ref{eq:idr22}) or (\ref{eq:idr24})  
into Eq. (\ref{eq:idr12}) 
 \begin{eqnarray}                       
F_q(\phi) &=& 
\frac{1}{2\pi z^q}\ \frac{z^2}{z^2+\alpha^2}, 
\label{eq:idr29}
\end{eqnarray}
where we have set $z= \exp(i\phi)$ and the subindex $q$ 
refers to the initial condition, $c_n(0) = \delta_{q,n}$. We note 
that $F_q(\phi)$ has poles at the origin and $z_p  =\pm i \alpha$. 
In terms of $z$, Eq. (\ref{eq:idr12}) becomes: 
\begin{equation}                       
c_k({\tilde  t}) = \frac{1}{2\pi i} \oint \ dz\ e^{i{\tilde t} (z+1/z)}\, 
\frac{u_k(z)}{z^{q-1}}\, \frac{1}{z^2+\alpha^2}, 
\label{eq:idr30}
\end{equation}
the path of integration being the unit circle. 
As shown by Longhi~\cite{Lo06}, Eqs. (9) to (12), to  extract the 
exponentially-decaying contribution to $c_k(t)$ one may use the residue 
theorem to evaluate 
the integral, retaining only the contribution from the Gamow pole at 
$z_g= -i\alpha$. The conjugate pole corresponds to exponential growth, which
is not the physical process, and the poles at the origin give compensating
contributions, leaving only a small remainder, at least in the survival 
probability. The Gamow pole contributes 
 
 \begin{equation}                       
c_{k,q}^{(e)}({\tilde  t}) 
= \frac{u_k(z_g)}{2 (-i\alpha)^{q}} \ e^{-\gamma {\tilde  t}/2}
\label{eq:idr31}
\end{equation}
where  the superscript (e) means exponential decay, and
 \begin{equation}                       
\frac{\gamma}{2} \equiv \frac{1 -\alpha^2}{\alpha}  = \frac{\Delta^2}{\alpha}.
\label{eq:idr32}
\end{equation}
This is a  remarkable result: it shows that all the initial 
conditions of the form $c_n(0)=\delta_{n,q}$ lead to the same 
lifetime, $\tau=1/\gamma$, for the exponentially decaying contribution to  
$|c_{k,q}({\tilde  t})|^2$, no matter which site $q$ is initially 
occupied. 

We develop the above result by evaluating $u_k(z_g)$. From Eq. 
(\ref{eq:idr11}), for $k \ge 2$, 
 \begin{eqnarray}                       
u_k(z_g) &=&  -~i^k\, \frac{1-\alpha^4}{\alpha^k},
\label{eq:idr33}
\end{eqnarray}
and therefore:
 \begin{eqnarray}                       
c_{k,q}^{(e)}({\tilde  t}) &=& -~i^{k+q}\ 
\frac{\Delta^2 (1+\alpha^2)}{2\ \alpha^{k+q}} 
\ e^{-\gamma {\tilde  t}/ 2}~, \quad k \ge 2.  
 \label{eq:idr34}
\end{eqnarray}
When $k=1$, the same calculation 
has a factor $1/\Delta$ in comparison: 
 \begin{equation}                       
u_1(z_g) = 
  -i\ \Delta\ \frac{1+\alpha^2}{\alpha}  ,
\label{eq:idr35}
\end{equation}
and 
 \begin{equation}                       
c_{1,q}^{(e)}({\tilde  t}) = - i^{q+1}\ \frac{\Delta(1+\alpha^2)}
{2 \alpha^{q+1}} \ e^{-\gamma {\tilde  t}/ 2}  . 
 \label{eq:idr36}
\end{equation}
The sudden approximation, Case 1, with $c_1(0)=1$,  
requires a separate discussion since $T_{2l}=0$ and $T_{2l+1} 
=(-)^2 \alpha^{2l}$, without the factor $\Delta$ of Eq. 
(\ref{eq:idr22}). One now finds 
\begin{equation}                       
F_1(\phi) = \frac{z}{2\pi \Delta(z^2 + \alpha^2)}
\label{eq:idr37}
\end{equation}
and
\begin{equation}                       
c_{k,1}^{(e)}({\tilde  t})= i^{k-1}\ 
\frac{\Delta(1+\alpha^2)}{2\alpha^{k+1}}  \ e^{-\gamma {\tilde  t}/ 2}
\label{eq:idr38}
\end{equation}
when $k \ge 2$, while
\begin{equation}                       
c_{1,1}^{(e)}({\tilde  t}) = \frac{1+\alpha^2}{2\alpha^2}  \ 
e^{-\gamma {\tilde  t}/ 2}.
\label{eq:idr39}
\end{equation}
This last result agrees with Eq. (11) of~\cite{Lo06}. 

Finally, we write the amplitudes for general initial 
occupations. Inserting the above results in Eq. (\ref{eq:idr06}) gives  
 \begin{eqnarray}                       
c_1^{(e)}({\tilde  t}) 
&=& \frac{1}{2i\alpha} (1+\alpha^2) \left(\sum_{q=1}^{\infty} {\tilde \vt}_q 
\frac{i^q}{ \alpha^q}\right)  \ e^{-\gamma {\tilde  t}/2}, \nonumber \\
c_k^{(e)}({\tilde  t}) 
&=& \frac{i^{k-2}}{2\alpha^k} \Delta (1+\alpha^2) \left(\sum_{q=1}^{\infty} 
{\tilde \vt}_q \frac{i^q}{ \alpha^q}\right)  \ e^{-\gamma {\tilde  t}/2}, 
\label{eq:idr40}
\end{eqnarray}
when $k \ge 2$. Clearly the exponential decay contribution has a  
delay $T$, manifest in the appearance of ${\tilde t} = t -T$ in these 
expressions. Furthermore, one sees that the decay constant $\gamma$ 
depends neither on the initial conditions nor on the site considered. 
\section{Initial switching}
We now address our main objective, which is time evolution when the 
coupling between site $1$ and site $2$, $\Delta_v(t)$, varies 
continuously with time. To do so we introduce a simple 
diagrammatic technique which describes visually the action of the time 
evolution operator. To introduce these diagrams we first apply 
them to the oft-studied case of ``sudden switching":  
$\Delta_v(t > 0 ) = \Delta= $constant, which will serve as a reference 
for results with variable $\Delta$.  

The time evolution operator can be written as 
\begin{eqnarray}                       
U(t) &=& \sum_{m=0}^{\infty} \   U_m(t) \qquad {\rm with} \nonumber \\
U_m(t) &=& (-i)^m \int_0^t \ dt_m \int_0^{t_m} \ dt_{m-1}
 \cdots \int_0^{t_2} \ dt_1 \nonumber \\
 &\times& H(t_m) H(t_{m-1}) \cdots H(t_1) 
 \label{eq:idr41}
\end{eqnarray}
and in the special case of a time independent $H$, this reduces to 
 $U(t) =$ exp$(-iHt) = \sum_{m=0}^{\infty} (-it H)^m/(m!)$. Suppose 
we start with the state $|1>$ with only site 1 occupied 
initially. Then 
\begin{eqnarray}                       
U(t) |1 \ra &=& |\Psi(t) \ra = \begin{pmatrix} c_1( t) \cr 
c_2( t) \cr \vdots \end{pmatrix} 
 \label{eq:idr42}
\end{eqnarray}

\subsection{Diagrammatic method and sudden switching}
 We now develop a diagrammatic method to calculate the time 
evolution operator, first in the context of sudden switching where the 
exact solution is already available. It will be extended to linear 
switching in the next subsection. To begin we compute $U_m(t)|1 \ra$ for 
increasing values of $m$, assuming $\Delta$ is constant. 
For  $m=0$, $U_0(t) = 1$, and therefore 
\begin{equation}                       
U_0(t) |1 \ra =|1 \ra. 
\label{eq:idr43}
\end{equation}
When $m=1$ 
 \begin{eqnarray}                       
U_1(t) |1 \ra &=& -i \int_0^t \ dt_1 \ H \ |1 \ra = -it H \ |1 \ra \nonumber \\
&=& i t \Delta\ |2 \ra .
\label{eq:idr44}
\end{eqnarray}
  \begin{figure} [htb]                         
\includegraphics[width=8cm]{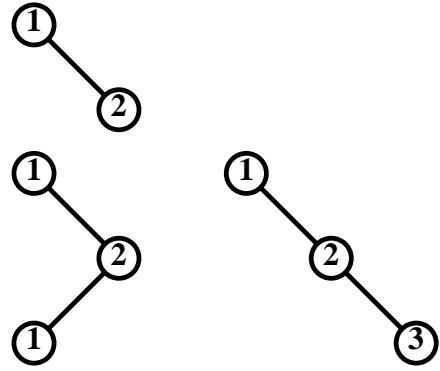}
\caption{Diagrams  which begin at site 1, with $m=1$ and $2$ lines.  
The meaning of the various symbols is explained in text.} 
 \label{gg1}
\end{figure} 
The first graph in Fig. \ref{gg1} describes this process: the circles 
represent the sites, the time evolution is downwards and the lines 
describe hopping between adjacent sites. The $m=2$ 
contribution is 
 \begin{eqnarray}                       
&&\,\, U_2(t) = (-i)^2 \int_0^t \ dt_2 \int_0^{t_2} \ dt_1 H^2 =
\frac{(-i)^2}{2!}t^2 H^2
\nonumber \\
&& U_2(t)|1 \ra = \frac{(-it)^2}{2 !} \left(\Delta^2|1 \ra 
+ \Delta|3 \ra\right) 
\label{eq:idr45}
\end{eqnarray}
and the two corresponding diagrams are in the lower row of Fig. \ref{gg1}. 
Continuation to higher $m$ is straightforward. The 
rules to generate the contributions to $U_m(t)|1 \ra$  are: 
1) each line joining $1$ and $2$ contributes a factor $-\Delta$, 
whereas the other lines contribute a factor $-1$ each; 2) if the 
diagram has $m$ lines: multiply by $(-it)^m/ (m!)$. Combining these 
two, the net result is to multiply by $(it)^m/(m!)$. The final state 
is the last vertex at bottom of the diagram. 
 \begin{figure} [htb]                         
\includegraphics[width=8cm]{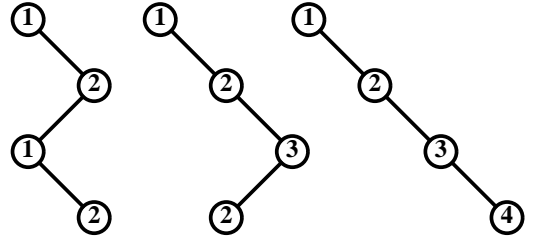}
\caption{Three line diagrams which begin at site 1.} 
 \label{gg2}
\end{figure} 
As a further example consider the $m=3$ contributions: 
applying the rules to the three graphs of Fig. \ref{gg2}, 
one can immediately write 
\begin{eqnarray}                       
U_3(t)|1 \ra &=& \frac{(it)^3}{3!} \bigg( \Delta^3 |2 \ra 
+ \Delta|2 \ra + \Delta|4 \ra \bigg).
\label{eq:idr46}
\end{eqnarray}
We now focus on contributions to $c_1(t)$, to all orders in $m$.
The relevant diagrams will be those  having  $|1 \ra$ 
as both initial and final state. Since this implies an even 
number of lines, we write $m=2l$ with $ l= 1, 2, \cdots$. 
  \begin{figure} [htb]                         
\includegraphics[width=8cm]{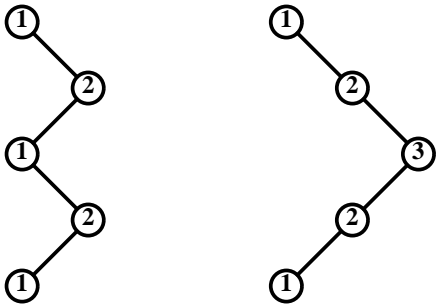}
\caption{Four line diagrams which begin and end at site $1 $. }
 \label{gg3}
\end{figure} 
The simplest of these diagrams, $l=1$, is already included in Fig. 
\ref{gg1}. The two diagrams with $l=2$ are shown in Fig. \ref{gg3}.  
In the special case $\Delta=1$, each  
diagram with the same number of lines gives an identical contribution to 
$c_1$. Applying the  general rules, the desired series is 
\begin{eqnarray}                       
c_1(t,\Delta=1) 
&=& \sum_{l=0}^{\infty} (-)^l \frac{t^{2l}}{(2l)!} \ N_{2l}^{1 \to 1},
\label{eq:idr47}
\end{eqnarray}
where $N_{2l}^{1 \to 1}$ is the number of distinct diagrams with $2l$ 
lines having $|1 \ra$ as both initial and final state. To determine
it, we rewrite Eq. (\ref{eq:idr17}) for the Case 1 initial conditions and  
$\Delta= 1$: it simplifies to   
 \begin{eqnarray}                       
c_1(t, \Delta=1 )  = \frac{J_1(2t)}{t} 
&=& \sum_{l=0}^{\infty}   \frac{(-)^l\, t^{2l}}{l! (l+1)!}, 
\label{eq:idr48}
\end{eqnarray}
using the Taylor series for $J_1(2t)$. Comparing Eqs. 
(\ref{eq:idr47}) and (\ref{eq:idr48}), it follows that 
\begin{equation}                       
N_{2l}^{1 \to 1}  = \frac{(2l)!}{l! (l+1)!}.
\label{eq:idr49}
\end{equation}
Now that the $N_l^{1 \to 1}$ are known, we can return to the more 
general case $\Delta\ne 1$ and consider those diagrams that give 
contributions of order $\Delta^2$ to $c_1(t)$. These begin and end 
at $|1 \ra$, but never have $|1 \ra$ as an intermediate state. 
The second diagram in Fig. \ref{gg3} is an example. 
It is now easy to write down their contribution: 
 \begin{eqnarray}                       
c_1^{(2)}\!\!&=&\!\! \Delta^2\! \sum_{l=1}^{\infty} \frac{(it)^{2l}}{(2l)!} \ 
N^{1 \to 1}_{2(l-1)}  
\!=\! \Delta^2 \sum_{l=1}^{\infty}\!  \frac{(-)^l \ t^{2l}} {(4l-2) (l!)^2}. 
\label{eq:exi15}
\end{eqnarray}
Here we have noted that the number of diagrams with $2(l-1)$ lines 
and beginning and ending at $|2 \ra$ without $|1 \ra$ as intermediate 
states, equals $N_{2(l-1)}^{1\to 1}$. Since we are interested in 
values of $\Delta \ll 1$, we will not write down 
expressions for contributions of order $\Delta^4$ or higher. This 
supposes that processes with  more than two tunneling events 
between sites 1 and 2 can be  neglected during the short initial-switching
time interval. 

To find the amplitude for site 2 we begin by considering again the 
special case $\Delta=1$. The simplest diagrams that connect $|1 \ra$ 
to $|2 \ra$ are shown in Fig. \ref{gg1} (in the upper row) and 
Fig. \ref{gg2} (the first two). They have an odd number of lines, 
$m=2l+1$, $l=0,1,2,\cdots$. Again all the diagrams with the same 
number of lines give equal contributions. Thus 
 \begin{equation}                       
c_2(t, \Delta=1) = \sum_{l=0} \frac{(it)^{2l+1}}{(2l+1)!} \ N_{2l+1}^{1
  \to 2}. 
\label{eq:idr51}
\end{equation}
Eq. (\ref{eq:idr17})  tells us that when $\Delta=1$ ($\alpha^2=0$) and Case 1
  initial conditions, 
 \begin{eqnarray}                       
c_2(t, \Delta=1) &=&  \frac{2i}{t} J_2(2t) 
 =  2 i t \sum_{l=0}  \frac{(-t^2)^{l} }{l! (l+2)!} 
\label{eq:idr52}
\end{eqnarray}
and therefore
 \begin{equation}                       
N_{2l+1}^{1 \to 2} = \frac{2(2l+1)!}{l! (l+2)!}.
\label{eq:idr53}
\end{equation}

We now return to the case $\Delta \ne 1$, and determine 
the contribution to $c_2$ linear in $\Delta$. The simplest 
diagrams are those in the top row of Fig. \ref{gg1}, and in Fig. 
\ref{gg2}. More generally: we include all diagrams that begin with a link from 
$1$ to $2$ and are completed with lines that go from $2$ to $2$ 
without having $1$  as an intermediate state. All such diagrams have 
an odd number $n=2l+1$ of lines, $ l =0, 1, \cdots$ and their contributions 
are 
 \begin{eqnarray}                       
c_2^{(1)}(t) &=& \Delta\sum_{l=0}^{\infty} \frac{(it)^{2l+1}}{(2l+1)!} \
N_{2l}^{1 \to 1} \nonumber \\ 
&=& i \Delta\sum_{l=0} (-)^l \frac{t^{2l+1}}{(2l+1)\ l!\ (l+1)! }, 
\label{eq:idr54}
\end{eqnarray}
where we have used that the number of the above sub-diagrams, beginning and
ending at $2$, equals $N_{2l}^{1 \to 1}$.\\
Finally, to determine the amplitudes for site $n$  we proceed similarly: 
For constant $\Delta=1$, 
 \begin{eqnarray}                       
c_n(t, \Delta=1) &=& \sum_{l=0}^{\infty} (-)^l
\frac{(it)^{n-1+2l}}{(n-1+2l)!} \ N_{n-1+2l}^{1 \to n} 
\label{eq:idr55}
\end{eqnarray}
and comparison to Eq. (\ref{eq:idr17}) with Case 1 initial conditions 
 \begin{eqnarray}                       
c_n(t, \Delta=1) &=& i^{n-1} \ \frac{n}{t} J_n(2t) \nonumber \\
&=& i^{n-1} n \sum_{l=0} (-)^l \frac{t^{n-1+2l}}{l!\, (n+l)!} 
\label{eq:idr56}
\end{eqnarray}
shows that
\begin{equation}                       
N_{n-1+2l}^{1 \to n} = \frac{n (n-1+2l)!}{l! (n+l)!},
\label{eq:idr57}
\end{equation}
which generalizes Eqs. (\ref{eq:idr49}) and (\ref{eq:idr53}). Finally, 
for constant $\Delta\ne 1$ we obtain the contribution to $c_n$ 
linear in $\Delta$ 
\begin{eqnarray}                       
c_n^{(1)}(t) &=& \Delta\sum_{l=0}^{\infty} 
\frac{n-1}{n-1+2l} \frac{(it)^{n-1+2l} }{ l! \,  (n-1+l)!}.
\label{eq:idr58}
\end{eqnarray}
\subsection{Linear switching}
When $\Delta(t)$ depends on time, the time evolution operator is 
given by Eq. (\ref{eq:idr41}) and 
the term of order $m$ involves $m$ consecutive integrations over 
intermediate times. In computing the contribution linear in $\Delta$ to 
$c_n(t)$, the time-dependent term will appear only in the first line 
of the diagram. Thus for $m=1$ 
 \begin{eqnarray}                       
U_1(t_2) |1 \ra &=& i \int_0^{t_2} \ dt_1 \ \Delta_v(t_1) \ |2 \ra 
\equiv f(t_2) |2 \ra 
\label{eq:idr59}
\end{eqnarray}
and we construct the contributions from the diagrams corresponding to 
higher orders by integrating $f(t)$ the required number of times. 
 \begin{equation}                       
{\cal I}_m (t) = \int_0^t \ dt_m \int_0^{t_{m-1}} 
\ dt_{m-1} \cdots\int_0^{t_3} \
dt_2 \ f(t_2)
\label{eq:idr60}
\end{equation} 
This ${\cal I}_m$ replaces the $t^m/m!$ found when $\Delta$ is 
constant. For the ``linear switching" case, Eq. (\ref{eq:idr02}) and $t 
<T$, the integral above gives (since $f(t)= t^2/2$), ${\cal I}_m 
= t^{m+1}/(m+1)!$. The modified rule for diagrams corresponding to 
linear switching is to replace $t^m/m!$ by $t^{m+1}/(m+1)!$ and 
also add the factor $1/T$ from Eq. (\ref{eq:idr02}).  Thus, instead of
Eq. (\ref{eq:idr58}),  taking  all relevant diagrams into account we find 
 \begin{eqnarray}                       
c_n^{(1)}(t) &=& \frac{\Delta}{T} \sum_{s=0}^{\infty} \frac{i^{n-1+2s}\ 
t^{n+2s}\ (n-1)}{(n+2s)(n-1+2s) s! \, (n-1+s)!} \nonumber \\ 
\label{eq:idr61}
\end{eqnarray}
for $t \le T$.  Note that the phase of this amplitude 
is $i^{n-1}$. This remains true if we include terms of higher order 
in $\Delta$. As discussed following 
Eq. (\ref{eq:idr45}), to connect site 1 with site $n$, the corresponding
diagrams must have  $n-1+ 2 s$ lines with $s= 1, 2, ...$. Therefore the
phase will be $i^{n-1+2s}$ and terms with odd (even) $n$ will be real 
(imaginary). 

\begin{figure} [htb]                         
\includegraphics[width=8cm]{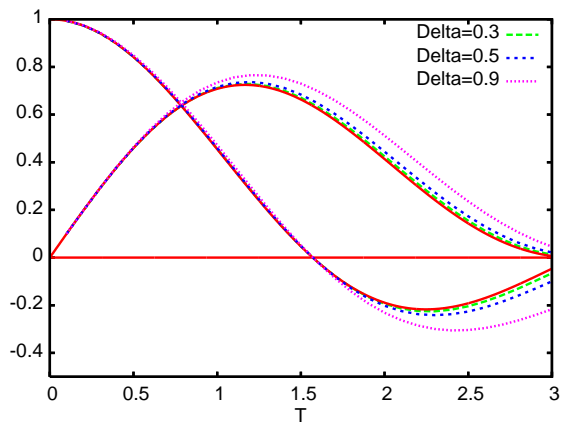}
\caption{(Color online)  Functions $S_{o}(T)/\Delta^2$ and 
Im($S_{e}(T)/\Delta^2$)
for the linear rise model with site 1 initially ($t=0$) 
occupied, and the rest empty. 
Various dashed lines: exact numerical calculations for $\Delta= 0.3, 0.5$ 
and $0.9$ according to key. Adjacent continuous lines: expressions of  
Eqs. (\ref{eq:idr64})  and (\ref{eq:idr65}).} 
 \label{emp2}
\end{figure}

Our next task is to apply these amplitudes to evaluate the sums
defined in Eqs. (\ref{eq:idr27}), to obtain approximations to the
asymptotic amplitudes:
\paragraph*{$S_o$ to lowest order in $\Delta$:} 
Since 
\begin{equation}                       
(1+\alpha^2)q -2\alpha^2 = 2(q-1) + \Delta^2(q-2) \simeq 2(q-1),
\label{eq:idr62}
\end{equation}
to first order in $\Delta$, therefore from Eq. (\ref{eq:idr27})               
\begin{equation}                       
S_o^{(1)}(T) = \Delta^2 + 4 \Delta \sum_{l=1}\ l \ c_{2l+1}^{(1)}(T).
\label{eq:idr63}
\end{equation}
Inserting the amplitudes  from Eq. (\ref{eq:idr61}), we find       
\begin{eqnarray}                       
S_o^{(1)}(T) 
&=& \Delta^2+ 4 \frac{\Delta^2}{T} \sum_{l=1}^{\infty} (-)^l \ l^2 \times 
\nonumber \\ 
&&   \sum_{s=0}^{\infty} (-)^s
\frac{T^{2(l+s)+1}}{(2(l+s)+1) (l+s) s! (2l+s)!}\nonumber \\ 
&=& \frac{\Delta^2}{2T} \sin 2T, 
\label{eq:idr64}
\end{eqnarray}
where the steps to perform the double sum are detailed in the Appendix.
\paragraph*{$S_e$ to lowest order in $\Delta$:}
The derivation is entirely similar to that of $S_o$, and leads to              
  
\begin{eqnarray}                      
S_e^{(1)}  &=& 2 \Delta \sum_{l=1}^{\infty}\ (2l-1)\ c_{2l}^{(1)}(T) \nonumber 
\\
&=& -i \frac{\Delta^2}{T} \sum_{l=1}^{\infty} (-)^l (2l-1)^2 
\sum_{s=0}^{\infty} (-)^s \times \nonumber \\
&& 
\frac{T^{2(l+s)}}{(l+s)(2(l+s)-1) s! (2l+s-1)! }\nonumber \\
&=& i \frac{\Delta^2}{T}
\sin^2 T,  
\label{eq:idr65}
\end{eqnarray}
where the sum is again detailed in the Appendix.
To picture how the functions $S_{o}$ and $S_{e}$ behave, we  
computed them by numerically solving the TDSE to determine the $\vt_q, q 
= 1, 2, \cdots$, and then computed $S_{o}$ and $S_{e}$ (also numerically)  
using Eqs. (\ref{eq:idr27}). Fig. \ref{emp2} shows the curves thus 
obtained for three values of $\Delta$. As expected, for the 
smallest $\Delta$ the curves are very close to the predictions of 
Eqs. (\ref{eq:idr64}) and (\ref{eq:idr65}), but even for 
$\Delta = 0.9$, the approximation is qualitatively quite 
satisfactory. The remaining difference has to be ascribed to 
contributions of higher order in $\Delta$. 
Finally, inserting Eqs. (\ref{eq:idr64}) and (\ref{eq:idr65}) into 
Eq. (\ref{eq:idr26}) we find that to lowest order in $\Delta$ for the $S_o(T)$
and $S_e(T)$, but keeping $\Delta$ to all orders in the later 
time-evolution,  
 \begin{eqnarray}                       
&& c_1^{(a)}( t) =  \frac{\Delta^2}{\sqrt{\pi} (t-T)^{3/2} (1+\alpha^2)^2} 
\nonumber \\
&\times& \quad \frac{\sin T}{T} \sin \left( 2(t-T)  -\frac{\pi}{4} + T\right). 
\label{eq:idr66}
\end{eqnarray}
The beauty of this expression is that it shows the effect of a finite 
rise-time $T$, very cleanly: 1) it adds a phase $T$ to the argument 
of the sine, and 2) it adds a factor $(\sin T)/ T$ to the amplitude 
of the sudden approximation (the $T \to 0$ limit.) One sees 
that for linear switching a finite $T$  always {\bf reduces} the 
amplitude. 
Repeating the same steps for the $c_n$, starting from 
Eq. (\ref{eq:idr20}), we arrive at
\begin{eqnarray}                     
&& c_n^{(a)}(t) = \frac{i^{n-1} [n +\alpha^2(n-2)] \Delta}{\sqrt{\pi} 
(t-T)^{3/2}  (1+\alpha^2)^2}  \times \nonumber \\ 
&& \quad \frac{\sin T}{T} \sin
\left(2(t-T)- (n- \half)\frac{\pi}{2}  +  T\right),  
\label{eq:idr67}
\end{eqnarray}
which shows the same memory effects. 
Previously, the reduction or enhancement of the amplitude had 
been discussed only for adiabatic switching: see Robinson,~\cite{Rob86} 
and references therein. Robinson studied the model of a discrete state coupled
to a continuum extending from a finite threshold to infinity. The explicit
forms of the continuum state density and of the adiabatic switching-on 
function 
for discrete to  continuum coupling, $f(t)$,  were left unspecified. Making 
only very general assumptions  he was able to show that the
asymptotic decay was controlled by the Fourier transform of 
the time-derivative ${\dot f}(t)$, 
but no results for specific attenuation functions were given. 
In this paper the continuum is that of the tight binding model;  it has 
finite width, and  Robinson's methods do not apply. 
By working with the Fano-Anderson model, we have obtained explicit forms, 
Eqs. (\ref{eq:idr66}) and (\ref{eq:idr67}), for the memory effects. 
\section{Results for linear switching}
\begin{figure} [htb]                         
\includegraphics[width=8cm]{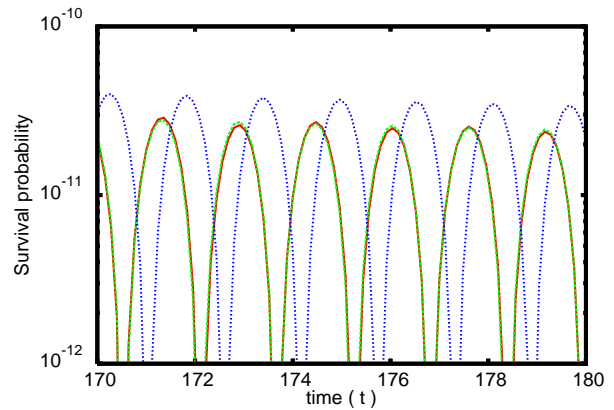}
\caption{(Color online) Survival probability in the Linear rise model, 
for $T=1$ and $\Delta=0.3$. 
The continuous (red) line  is the exact result. Practically hidden by it, 
the (green) dashed line is the value from the asymptotic 
expressions in the text. For comparison, the sudden 
approximation is also shown, as a (blue) dotted line.} 
 \label{inias1}
\end{figure}

\begin{figure} [htb]                         
\includegraphics[width=8cm]{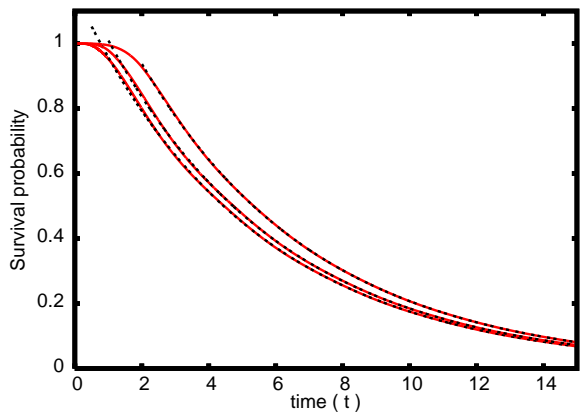}
\caption{(Color online) Linear rise model, $\Delta=0.3$: 
Survival probabilities for $T=0.5, 1$ and $2$ from left to right. 
Continuous (red) lines: exact. 
Dotted (blue) lines: the exponential decay component, Eq. (\ref{eq:idr40}).} 
 \label{iniex1}
\end{figure}

\paragraph*{Amplitudes at $t=T$.}
To illustrate our results we choose $\Delta=0.3$ as in Fig.1. As has 
been shown, Eq. (\ref{eq:idr32}), the lifetime is $\tau = 1/\gamma 
= 5.3$. We only consider values of the rise time $T \ll \tau$; 
otherwise the rise and decay regimes would be mixed.  We first 
computed the occupation amplitudes, $c_n(T)$, by 
numerical solution of the time dependent Schr\"odinger equation as 
described in~\cite{Koo90}.  These were compared to the 
approximate analytical expressions, eq. 
(\ref{eq:idr61}), derived in Section III. The agreement with the 
exact values of $c_n(T)$ is excellent, moreso for 
small values of $T$. But even at $T=2$ the absolute values are 
very close: in the format  ``{\rm exact (approximation)}", we have for 
$n=2$: $ 0.222\, (0.219)$, $n=3$: $0.136\, (0.136)$, $n=4$: $0.0676\, 
(0.0675)$,
$n=5$: $0.0275\, (0.0274)$.  
\paragraph*{Asymptotic survival probabilities.}
To compute the asymptotic amplitude $c_1^{(a)}({\tilde t})$. we then use  
Eqs. (\ref{eq:idr26}), 
(\ref{eq:idr27}) and the above exact values for the  amplitudes at $t=T$.
This is compared to the exact numerical  solution 
of the TDSE  in Fig. \ref{inias1}, for the case of rise  time $T=1$. 
Qualitatively similar oscillations are found for other $T$, even as 
high as $T=2$. In all cases the prediction fom Eq. (\ref{eq:idr28}) 
is excellent. Note also that there is a clear difference from the 
prediction given by the ``sudden approximation" (rise time $T=0$.) We 
stress that the envelope of the oscillations in $|c_1^{(a)}(t)|^2$ 
follows the predicted $t^{-3}$ algebraic law and that the period of 
the oscillations, according to Eq. (\ref{eq:idr26}) or 
(\ref{eq:idr28}), is $ \pi$. Neither result is affected by the finite 
rise time $T$; only the envelope and  phase of the oscillations are 
affected by the initial-switching process. Finally, we have also plotted the
prediction from the approximate expression, eq. \ref{eq:idr66}. On the scale
of the figure it cannot be distinguished from the green line previously
obtained, thus confirming the accuracy of our approximate analytic expressions.
Similarly, we have  tested the 
accuracy  of Eq. (\ref{eq:idr67}) by comparison to exact TDSE calculations:   
The agreement  found for the other $c_n^{(a)}(t)$ is similar to that for the
amplitudes of the first site. Further, we 
have found that the additional phase $T$ in the argument of the sine, 
in Eq. (\ref{eq:idr66}) has the same value whatever the shape of the 
switching function, providing that the average value of $\Delta_v(t)$ 
is half of the ultimate value $\Delta$, over the interval $0<t<T$. 
Finally, we have explored the accuracy of the ansatz $c_n(t) =
c_n^{(e)}(t) + c_n^{(a)}(t)$ for $n=1, 2, ...$:  For $n=1$ and the range of
times of Fig.  1, the exact (numerical) result and the values from 
the ansatz are indistinguishable on the scale of that figure. 
Similar agreement is found for other small
values of $n$. Support for the validity of this ansatz, for the sudden
approximation, can be found in the approach based on Laplace transforms
presented in \cite{Lo06c}: the asymptotic term is an approximation to
the contribution due to the Bromwich paths, and to this should be added that
of the poles, which produce the exponential decay. 
\paragraph*{Exponential decay regime}
In Fig. \ref{iniex1} we compare the exact 
survival probability $|c_1|^2$, again determined from a numerical solution 
of the TDSE, to the predictions from Eq. (\ref{eq:idr40}). Three 
representative values of $T$ are shown. Except at small times $t \le T$, the 
exponential contribution dominates and accurately follows the 
behaviour of the survival probability. Even for values of $T$ well 
beyond those suggested by the ``sudden approximation'' condition ($ T 
\Delta  \ll 1$)~\cite{Me58} one sees that the decay is exponential 
with the same decay constant.   
 \begin{figure} [htb]                         
\includegraphics[width=8cm]{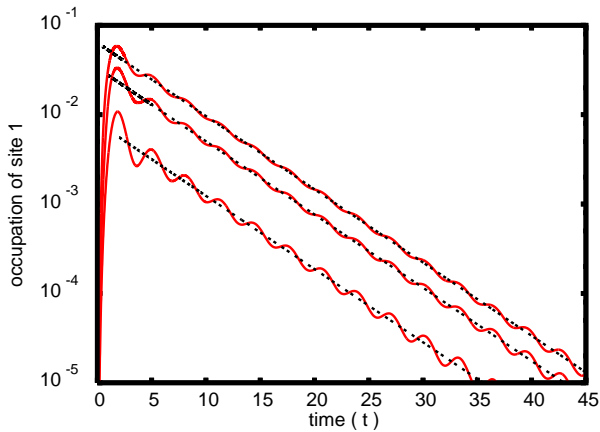}
\caption{(Color online)  $|c_1(t)|^2$ for, from top to bottom, rise times
 $T=0.5, 1$ and $2$. Linear rise model with $\Delta=0.3$, but now 
site 2 is initially occupied, and the  others empty.   Continuous 
(red) lines: exact. Dotted (blue) straight lines: the exponential decay 
component, Eq. (\ref{eq:idr40}).} 
 \label{iniex3}
\end{figure} 

 In Figs. \ref{inias1} and \ref{iniex1} we assumed that at $t=0$ only 
the first site was occupied. But, as already stressed, Eqs. 
(\ref{eq:idr40}) are valid for any set of occupation amplitudes at 
${\tilde t}=0$. As a second example, we explore what occurs when only 
site 2 is occupied at $ t=0$. Fig. \ref{iniex3} shows the 
occupation  probability $|c_1|^2$ for the linear rise model and 
several values of $T$. Exact (numerical) solutions of the TDSE are 
compared to values from  Eq. (\ref{eq:idr40}). (The relevant set of 
$\vt_q, q=1,2, ..$ was generated  by solving the TDSE using 
$\Delta_\ell(t)$.) The exponentially decaying contribution is dominant, 
except at very small times during which the amplitude $c_1(t)$ builds 
up. The maximum occupation is of order $\Delta^2$, and is lower for 
larger $T$, because site 2 is decaying outwards during the 
switching-on process before site 1 can begin to fill. Superimposed on the 
exponential there are pronounced oscillations, due to terms neglected 
when only the pole at $z_g = -i \alpha$ was retained. The average 
trend is well explained by the exponential. 

We have made similar comparisons when initially only site $n=5, 10$ 
or $20$ is occupied. In all cases the trend is similar: there is an 
initial period where the occupation of site 1 accumulates and its 
subsequent exponential decay is well described by Eq. (\ref{eq:idr40}). 
Increasing the value of $\Delta$ increases the decay constant, so the 
changeover from exponential to the asymptotic decay regime occurs 
sooner. 

\section{Summary and Discussion}
In the context of Longhi's version of the Fano-Anderson model we have 
obtained explicit expressions for  so-called ``memory" effects.  In 
other models these can be studied only numerically or discussed in 
very general terms, leading to few detailed predictions. The 
switching-on process occurs during a time interval $T$ between the 
original preparation of the system in site 1, and the stabilization 
of the coupling to the external sites at its final value $g \Delta$. 
We have shown that no matter what profile $\Delta_v(t)$ is chosen, 
the site amplitudes oscillate asymptotically with period $\pi$, with 
an envelope having its leading term proportional to $t^{-3/2}$. 
Furthermore, we have also shown that when exponential decay occurs, 
the decay constant is the same for all site amplitudes.

For the special case of linear switching, we have derived 
series expansions for the contributions linear in $\Delta$ to the 
site amplitudes at time $T$, Eq. (\ref{eq:idr61}). We have 
found that to this order the 
asymptotic amplitude of site 1 is given by Eq. (\ref{eq:idr66}), 
which shows in a very clean way that memory effects introduce a 
delay $T$ in the phase, and that the modulation of the amplitude is 
given by a factor $(\sin T)/T$.  
Combined with the exponentially decaying contribution they allow  
accurate predictions in the intermediate range of times where 
interference between the two contributions occurs, and this opens the 
way to examine memory effects in this range of times.

The methods of Section III, could be applied to quadratic or other forms of 
initial switching, allowing a systematic study of memory effects. Our 
predictions for the occupation amplitudes in the initial stages 
should be amenable to experimental verification using the waveguide 
analogue of the tight-binding model. 

\section{Acknowledgements}
We are grateful to DGES-Spain for support through grant  
FIS2006-10268-C03-01;   
to UPV-EHU for grant (GIU07/40); and to  NSERC-Canada for 
Discovery grants RGPIN-3198 (DWLS), SAPIN-8672 (WvD). 

\appendix 
\section{}
To sum the double series in Eq. (\ref{eq:idr64}) we make a 
change of index to $k=l+s$         
 \begin{eqnarray}                       
D  &\equiv& \sum_{l=1} (-)^l l^2 \ \sum_{s=0} (-)^s
\frac{T^{2(l+s)+1}}{(2(l+s)+1) (l+s) s! (2l+s)!} \nonumber \\
&=& \sum_{k=1}^{\infty} (-)^k \frac{T^{2k+1}}{k(2k+1)} \ \sum_{s=0}^k
\frac{(k-s)^2}{s! (2k-s)!} ,
\label{eq:idra01}
\end{eqnarray}
and to perform the sum over $s$, we write 
 \begin{eqnarray}                       
\sum_{s=0}^k
\frac{(k-s)^2}{s! (2k-s)!} &=& \sum_{s=0}^{2k}
\frac{(k-s)^2}{s! (2k-s)!} - \sum_{s=k+1}^{2k}
\frac{(k-s)^2}{s! (2k-s)!} \nonumber \\
&=& \sum_{s=0}^{2k}
\frac{(k-s)^2}{s! (2k-s)!} - \sum_{s'=0}^{k-1} \frac{(k-2k+s')^2}{(2k-s')!
  (s')! } \nonumber \\
&=& \sum_{s=0}^{2k}
\frac{(k-s)^2}{s! (2k-s)!} - \sum_{s=0}^{k-1}
\frac{(k-s)^2}{s! (2k-s)!},  
\label{eq:idra02}
\end{eqnarray}
where we have introduced $s'=2k-s$ to rewrite the second sum. Next, noticing
that
 \begin{equation}                       
\sum_{s=0}^k \frac{(k-s)^2}{s! (2k-s)!} = \sum_{s=0}^{k-1}
\frac{(k-s)^2}{s! (2k-s)!} . 
\label{eq:idra03}
\end{equation}
we finally arrive at
 \begin{equation}                       
\sum_{s=0}^k
\frac{(k-s)^2}{s! (2k-s)!} = \frac{1}{2} \sum_{s=0}^{2k}
\frac{(k-s)^2}{s! (2k-s)!}.
\label{eq:idra04}
\end{equation}
Using the identities
 \begin{eqnarray}                       
\sum_{s=0}^{2k} \begin{pmatrix} 2k \cr s \end{pmatrix} &=& 2^{2k} \nonumber \\
\sum_{s=0}^{2k} s \begin{pmatrix} 2k \cr s \end{pmatrix} &=& 2^{2k}\ k  
\nonumber \\  
\sum_{s=0}^{2k} s(s-1) \begin{pmatrix} 2k \cr s \end{pmatrix} &=& 2^{2k-1} k
\left( 2k - 1 \right)  
\label{eq:idra05}
\end{eqnarray}
and decomposing the numerator of Eq. (\ref{eq:idra03}) as $(k-s)^2= 
k^2-(2k-1)s+
s(s-1)$, Eq. (\ref{eq:idr64})  simplifies to         
 \begin{eqnarray}                       
S_o^{(1)}(T) &=&   \frac{\Delta^2}{2T} \sum_{k=0} (-)^k
\frac{(2T)^{2k+1}}{(2k+1)!} 
= \Delta^2\, \frac{\sin 2T}{2T}~.~~ 
\label{eq:idra06}
\end{eqnarray}

The double series in Eq. (\ref{eq:idr65}) is summed using the same methods:
We again write $k=l+s$ to replace the sums over $l$ and $s$ by 
sums over $k$ and $s$:        
 \begin{eqnarray}                       
S_e^{(1)} &=& -i \frac{\Delta^2}{T} \sum_{k=1}^{\infty} \frac{(-)^k
  T^{2k}}{k(2k-1)} \sum_{s=0}^{k-1} \frac{(2k-2s-1)^2}{s!(2k-s-1)!}, 
\nonumber \\ 
\label{eq:idra07}
\end{eqnarray}
and to evaluate the sum over $s$ we proceed as before 
 \begin{eqnarray}                       
& & \sum_{s=0}^{k-1} \frac{(2k-1-2s)^2}{s! (2k-1-s)!} \nonumber \\ 
&=& \sum_{s=0}^{2k-1} \frac{(2k-1-2s)^2}{s! (2k-1-s)!}  - 
\sum_{s=k}^{2k-1} \frac{(2k-1-2s)^2}{s! (2k-1-s)!} \nonumber \\
&=& \sum_{s=0}^{2k-1} \frac{(2k-1-2s)^2}{s! (2k-1-s)!} - \sum_{s'=0}^{k-1}
\frac{(2k-1-2(2k-1-s'))^2}{(s')! (2k-1-s')!} \nonumber \\
&=& \sum_{s=0}^{2k-1} \frac{(2k-1-2s)^2}{s! (2k-1-s)!} - \sum_{s=0}^{k-1}
\frac{(2k-1-2s)^2}{s! (2k-1-s)!} 
\label{eq:idra08}
\end{eqnarray}
and therefore
 \begin{equation}                       
\sum_{s=0}^{k-1} \frac{(2k-1-2s)^2}{s! (2k-1-s)!}  = \frac{1}{2}
\sum_{s=0}^{2k-1} \frac{(2k-1-2s)^2}{s! (2k-1-s)!}. 
\label{eq:idra09}
\end{equation}
Next we decompose the numerator as 
 \begin{equation}                       
(2k-1-2s)^2 = (2k-1)^2 -8(k-1) s + 4 s(s-1) 
\label{eq:idra10}
\end{equation}
and use Eq. (\ref{eq:idra05}), with replacement $2k \to 2k-1$
\ignore{
and use the combinatorial identities:
 \begin{eqnarray}                       
\sum_{s=0}^{2k-1} \begin{pmatrix} 2k-1 \cr s \end{pmatrix} &=& 2^{2k-1}
\nonumber \\
\sum_{s=0}^{2k-1} \ s \ \begin{pmatrix} 2k-1 \cr s \end{pmatrix} 
&=& (2k-1) 2^{2k-2} \nonumber \\
\sum_{s=0}^{2k-1} \ s (s-1) \ \begin{pmatrix} 2k-1 \cr s \end{pmatrix} 
&=& (2k-1) ( 2k-2)  2^{2k-3}
\nonumber \\
\label{eq:idra11}
\end{eqnarray}
}  
to rewrite the sum as
 \begin{eqnarray}                       
& &\sum_{s=0}^{k-1} \frac{(2k-1-2s)^2}{s! (2k-1-s)!}  
= \frac {2^{2k-2}}{(2k-2)!}.
\label{eq:idra12}
\end{eqnarray}
We now insert this in Eq. (\ref{eq:idra07}) and find       
 \begin{eqnarray}                       
S_e^{(1)} 
&=& i \frac{\Delta^2}{2T} \sum_{k=1}^{\infty} (-)^k \frac{(2T)^{2k}}{(2k)!}
\nonumber \\ &=& i \frac{\Delta^2}{2T} (1- \cos 2T ) = i \frac{\Delta^2}{T}
\sin^2 T .
\label{eq:idra13}
\end{eqnarray} 



\vfill\noindent file: initex.tex   May 11, 2009 

\end{document}